\documentclass[aps,10pt, prl,twocolumn,superscriptaddress]{revtex4-1}
\usepackage{graphicx}
\usepackage{float}
\usepackage{amsmath}
\usepackage{mathtools} 
\usepackage{color}
\usepackage{afterpage}
\usepackage{xr}
\usepackage{makecell}

\usepackage{amssymb}
\usepackage{graphicx,color} 
\usepackage{bbm} 
\usepackage{multirow}
\usepackage{enumitem}
\usepackage{mathrsfs}
\usepackage{commath}
\usepackage{hyperref}
\usepackage{fontenc}
\usepackage[dvipsnames]{xcolor}
\hypersetup{colorlinks=true,urlcolor=blue,citecolor=blue,linkcolor=blue}
\usepackage{tabularx}

\usepackage[english]{babel}
\usepackage{float}
\usepackage{amsmath,amssymb}
\usepackage{mathtools} 
\usepackage{color}
\usepackage{xr}
\usepackage{graphicx}
\usepackage{dcolumn}
\usepackage{bm}

\newcolumntype{Y}{>{\centering\arraybackslash}X}

\begin{document}

\preprint{APS/123-QED}

\title{Comment on ``Explicit Analytical Solution for Random Close Packing in $d = 2$ and $d = 3$''}

\author{Patrick Charbonneau}
\affiliation{Department of Chemistry, Duke University, Durham, North Carolina 27708}
\affiliation{Department of Physics, Duke University, Durham, North Carolina 27708}
\author{Peter K. Morse}
\thanks{Corresponding author}
\email{peter.k.morse@gmail.com}
\affiliation{Department of Chemistry, Duke University, Durham, North Carolina 27708}

\date{\today}

\maketitle

A recent letter proposes a first-principle computation of the random close packing (RCP) density in spatial dimensions $d=2$ and $d=3$~\cite{zaccone_explicit_2022}. This problem has a long history of such proposals~\cite{williams_packing_1998,jalali_estimate_2004,kamien_why_2007, song_phase_2008, jin_application_2010}, but none capture the full picture. Reference~\cite{zaccone_explicit_2022}, in particular, once generalized to all $d$ fails to describe the known behavior of jammed systems in $d>4$, thus suggesting that the low-dimensional agreement is largely fortuitous.

The crux of the proposed scheme is to evaluate the radial distribution function $g(r,\widehat{\varphi})$ at interparticle contact $r=\sigma^+$ for a (scaled) volume fraction ${\widehat{\varphi} = 2^d\varphi/d}$~\cite{parisi_mean-field_2010}, and to construct an invariant ${g_0= z\sigma/[d^2\widehat{\varphi}g(\sigma^+,\widehat{\varphi})]}$ for $z$ kissing contacts. Given the functional form $g(r,\widehat{\varphi})$, then $\widehat{\varphi}$ and $z$ can be evaluated at a known point to obtain an implicit expression, $\widehat{\varphi}(z)$, that holds for other packings. The scheme uses crystal close packings (CCP), for which $\widehat{\varphi}_\mathrm{CCP}$ and $z_\mathrm{CCP}$ are known, as reference. It is then possible to solve for the marginal-stability condition, $z_\mathrm{RCP}=2d$, which holds at jamming, to determine $\widehat{\varphi}_\mathrm{RCP}$. The only caveat identified in Ref.~\cite{zaccone_explicit_2022} is that one must choose a liquid structure expression that is defined from RCP to CCP. The two options suggested---the Carnahan-Starling (CS) equation of state~\cite{song_why_1989,charbonneau_glass_2011} and the Percus-Yevick (PY) closure relation---both fairly accurately describe the $d=3$ liquid structure. (Another caveat is that these liquid state descriptions do not capture singular features of $g(r,\widehat{\varphi})$ at RCP~\cite{charbonneau_glass_2017}.)

\begin{figure}[htp]
\includegraphics[width=\linewidth]{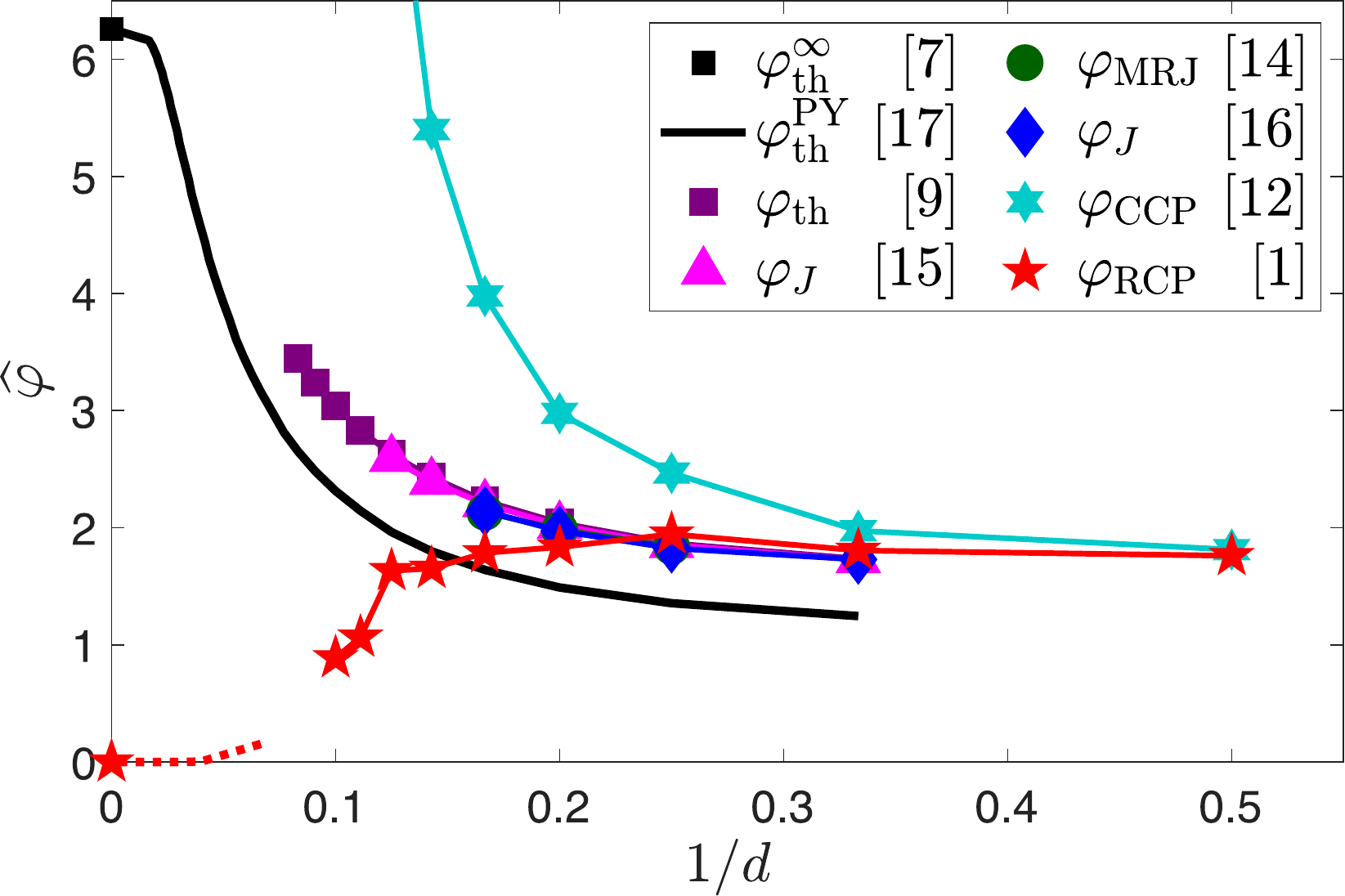}
\caption{Comparison of $\widehat{\varphi}_\mathrm{RCP}$ (red, dashed for high-$d$ asymptotic) with different numerical results for jamming (lines are guides to the eye).
Despite marked differences between the computational schemes, all trend similarly to the jamming threshold estimated using PY, $\widehat{\varphi}_\mathrm{th}^\mathrm{PY}$~\cite{mangeat_quantitative_2016} (black), which converges to the $d\rightarrow\infty$ prediction~\cite{parisi_mean-field_2010}. Note that the densest crystal density, $\widehat{\varphi}_\mathrm{CCP}$~\cite{conway_sphere_1998} (turquoise), grows increasingly distant from jamming with $d$.} \label{fig:phiJCompare}
\end{figure}

Because PY fails to describe essential contact features of $g(r,\widehat{\varphi})$ in $d>3$ liquids~\cite{charbonneau_dimensional_2021}, the dimensional generalization here focuses on the CS form. Its one parameter, $A(d)$, is known numerically for $d=2$-$13$, and asymptotically scales as $A(d) \sim 3^{(d+1)/2}\sqrt{2/(d\pi)}$~\cite{charbonneau_dimensional_2021}. Equating the invariant in both the RCP and crystalline phases yields 
\begin{equation}
\frac{\varphi_\mathrm{CCP}}{z_\mathrm{CCP}} \frac{1+A(d) \varphi_\mathrm{CCP}}{(1-\varphi_\mathrm{CCP})^d} = \frac{\varphi_\mathrm{RCP}}{2d} \frac{1+A(d)\varphi_\mathrm{RCP}}{(1-\varphi_\mathrm{RCP})^d},
\end{equation}
from which $\widehat{\varphi}_\mathrm{RCP}$ can be straightforwardly obtained in $d=2-10$ using CCP properties~\cite{conway_sphere_1998, charbonneau_thermodynamic_2021}. Because the method is independent of the crystal chosen, we also consider $D_d$ (generalized FCC) lattices, with ${\widehat{\varphi} = 2^{d/2-1}\pi^{d/2}/[d\Gamma(d/2+1)]}$ and ${z = 2d(d-1)}$~\cite{conway_sphere_1998}, to obtain the high $d$ asymptotic scaling ${\widehat{\varphi}_\mathrm{RCP} \sim (2\sqrt{\pi})^{d-1}e^{d/2}d^{-(5+d)/2}}\rightarrow0$. Comparing either of these results with RCP determined using a variety of numerical methods~\cite{skoge_packing_2006,charbonneau_glass_2011, morse_geometric_2014,charbonneau_memory_2021} shows a marked discrepancy for $d>4$ (Fig.~\ref{fig:phiJCompare}). In principle, $\widehat{\varphi}_\mathrm{RCP}$ should trend as these results, but its scaling is nowhere near them. The quantitative proximity in $d=2$-$4$ must therefore be considered accidental.

The key weakness of the approach of Ref.~\cite{zaccone_explicit_2022} is that it uses liquid state forms outside of their domain of validity. CS and PY capture the equilibrated liquid branch, but not the crystal branch ending at $\widehat{\varphi}_\mathrm{CCP}$ nor the out-of-equilibrium fluid branch terminating at $\widehat{\varphi}_\mathrm{RCP}$. (See Refs.~\cite{mangeat_quantitative_2016, parisi_mean-field_2010} for a careful discussion.) In low $d$, the relative proximity between $\widehat{\varphi}_\mathrm{RCP}$ and $\widehat{\varphi}_\mathrm{CCP}$ results in relatively small deviations from either form, but as $d$ increases these quantities separate significantly, thus revealing more pointedly the physical inadequacy of the proposed scheme.  By contrast, the approach of Ref.~\cite{mangeat_quantitative_2016}, which treats metastability more carefully, properly captures the dimensional trend, and could in principle be quantitatively improved to systematically close the gap with low-dimensional experiments and simulations.

\paragraph{Acknowledgments --}
We thank Ludovic Berthier, Eric Corwin and Francesco Zamponi for helpful discussions. This work has received funding from the Simons Foundation (Grant No. 454937).

\bibliography{highDStruct}

\end{document}